\documentclass[pdftex,aps,pre,amsmath,amsfonts,amssymb,superscriptaddress,noshowpacs]{revtex4}
\usepackage{graphicx}
\usepackage{color}
\usepackage{epsfig}
\usepackage[]{amsfonts}
\newcommand{\bea}{\begin{eqnarray}}
\newcommand{\misfit}{\mu}

\newcommand{\bsa}{\begin{subeqnarray}}
\newcommand{\esa}{\end{subeqnarray}}

\newcommand{\xtext}[1]{\mbox{#1}}

\begin{document}

\author{Arvind Baskaran}
\affiliation{Department of Mathematics, University of California, Irvine, CA 92697-3875.}
\email{baskaran@math.uci.edu}
\author{ Christian Ratsch}
\affiliation{Department of Mathematics and Institute for Pure and Applied Mathematics, University of California, Los Angeles, CA}
\email{cratsch@ipam.ucla.edu}
%\altaffiliation[Also at : ]{Michigan Center for Theoretical Physics, University of Michigan, Ann Arbor, Michigan 48109.}
\author{Peter Smereka\footnote{Deceased}}
\affiliation{Department of Mathematics and Michigan Center for Theoretical Physics, University of Michigan, Ann Arbor, MI 48109.}
\email{psmereka@umich.edu}

\title{Modeling the Elastic Energy of Alloys: Potential Pitfalls of Continuum Treatments}

\begin{abstract}
Some issues that arise when modeling elastic energy for binary alloys are discussed
within the context of a Keating model and density functional calculations.
The Keating model is a simplified atomistic formulation based on modeling
elastic interactions of a binary alloy with harmonic springs whose equilibrium
length is species dependent. It is demonstrated that
the continuum limit for the strain field are the usual equations of linear elasticity for alloys
and that they correctly capture the coarse-grained behavior of the displacement field.
In addition, it is established that Euler-Lagrange equation of the continuum limit
of the elastic energy will yield the same strain field equation. This is  the same
energy functional that is often used to model elastic effects in binary alloys.
However, a direct calculation of the elastic energy atomistic model reveals
that the continuum expression for the elastic energy is both qualitatively
and quantitatively incorrect.  This is because it does not take atomistic scale
compositional non-uniformity into account.  Importantly, this result also shows that finely mixed alloys
tend to have more elastic energy than segregated systems, which is the exact opposite of
predictions made by some continuum theories. It is also shown that for strained thin films
the traditionally used effective misfit for alloys systematically
underestimate the strain energy. In some models, this drawback is handled by
including an elastic contribution to the enthalpy of mixing which is characterized
in terms of the continuum concentration. The direct calculation of the atomistic
model reveals that this approach suffers serious difficulties. It is demonstrated
that elastic contribution to the enthalpy of mixing is non-isotropic and scale
dependent. It is also shown that such effects are present in density-functional theory
calculations for the Si/Ge system.
This work demonstrates that it is critical to include the microscopic arrangements
in any elastic model to achieve even qualitatively correct behavior.
\end{abstract}
\maketitle
\section{Introduction}

Many modern material systems consist of alloys of two or more species.
Applications of such alloy systems include semiconductor systems for new optoelectronic devices~\cite{Shchukin_1999},
oxides for optical and electronic applications~\cite{Hwang_2012}, hydrogen storage systems~\cite{Yang_2009}, and more.
Different species typically have different lattice constants leading to elastic strain, which can significantly impact
the performance and stability. Of particular interest is the strain driven formation and self-organization of
heterostructures such as quantum dots for semiconductor systems~\cite{Mo_1990, Spencer_2001, Tu_2004,MHS}.
It is therefore of paramount importance to develop models that properly describe the effect of strain in alloy systems.

The challenge in modeling alloys is that the total energy of an alloy system depends on
the composition.
This compositional dependence is hard to characterize and a common approach
is to assume a species dependent bond lengths and bond energies.
This allows one to separate the total energy into the chemical and elastic parts.
The alternative is to work with just the total energy.
This is usually not tractable and the separation allows one to construct
tractable models.
In the case of alloys of lattice mismatched elements, the difference in lattice spacing
introduces compositionally dependent strain.
This is not simply characterized by modeling the response of the material to applied stress.
These alloys retain a stress free strain.
This strain is determined as the deviation of the bond lengths from their equilibrium values.
However, if the bond lengths in the bulk are environmentally dependent, it would not be possible to
decompose the energy into chemical and elastic parts.
Fortunately an environmentally independent bond length appears to be a reasonable approach
to modeling microscopic strain in alloys (see Tsao \cite{TSAO} page 94).

In this paper we revisit the issue of modeling of elastic energy of a binary alloy.
The models for elastic energy of alloys fall under two broad categories, namely
continuum models and discrete/atomistic models.  Reality being that alloys are made of
discrete atoms, a fully atomistic description would be
the model of choice.  However, for practical reasons continuum models are often preferred.

Continuum models aim to characterize the alloy  in terms of macroscopic quantities.
These macroscopic quantities must vary slowly on the atomistic scale for a continuum model to be
consistent.  For an alloy, the concentration field does not in general vary slowly on the atomistic scale.
However, one can introduce an average concentration which can be constructed to vary smoothly
on atomistic scales.
It is apparent that for alloys one can make good predictions
of coarse-grained displacement fields using continuum theory.

However, the issue of deducing the elastic energy using continuum theory is more difficult.
A common starting point is to choose the reference lattice for the alloy in accordance
to Vegard's law based on the average concentration.
Vegard's law states that the average lattice spacing of an alloy varies linearly with concentration  (see \cite{TSAO}).
A second but more important assumption is that the stress free strain of a uniform alloy in this
reference state is zero.
Under these assumptions, it can be shown \cite{CAHN,FPL} that for a nonuniform isotropic alloy
the elastic energy can be written as
\begin{equation}
W_C  = \frac{\eta^2 E}{1-\nu}\int_V ( \theta - \theta_0 )^2 dV.
\label{eq:CAHN}
\end{equation}
In Eq.~(\ref{eq:CAHN}), $\eta$  is the effective misfit, $E$ is Young's modulus, $\nu$ is the Poisson ratio,
$\theta$ is the composition field of the alloy and $\theta_0$ is its average.
This formulation has been used to study a wide range of problems including
strain driven morphological instabilities,
spinodal decomposition, segregation and microstructure evolution in metal and
semiconductor alloys~\cite{FPL, Spencer_2001, Tu_2004,MHS,LONG_CHEN}. The formula given by
Eq. (\ref{eq:CAHN}) implies that non-uniformity of the alloy will result in an increase in
elastic energy. For this reason, it is often suggested that contributions to the free
energy that arise from atomistic misfit  stabilize an alloy or equivalently
 that intermixing will lower the elastic energy of the alloy.

  While Vegard's law in itself is known to be true for some systems \cite{VL1,VL2},
 it is a statement about the average lattice spacing of alloys. In particular it does not
 exclude the existence of microscopic stress free strain with respect to the reference lattice.
Using a Keating model it has been argued (e.g. Tsao \cite{TSAO}) that for alloys there
is microscopic strain even when the alloy atoms are placed in a lattice given by Vegard's law.
Thus, the stress free strain is not zero even when the composition appears to be macroscopically
uniform.  The energy stored in the springs has been used
as an estimate for the contribution  to the elastic energy due to this microscopic strain
(e.g. Tsao \cite{TSAO}).
The strain energy associated with this microscopic strain can be considered the elastic contribution
to the enthalpy of mixing, $H$.
Then one could posit (e.g. de Fontaine \cite{DEF} or Ren et al \cite{RWST}) that the
total elastic energy of the alloy can be written as
\begin{equation}
W=W_C+H.\label{MME}
\end{equation}
The atomistic scales are captured by $H$ whereas $W_C$ is used for the continuum scales.
This is only true in the presence of a clear separation of scales.
In many time dependent problems there is a range of scales, especially as the system coarsens.
For example, in a finely mixed alloy the dominant contribution will come from $H$, while the microscopic strain energy
will lessen in favor of $W_C$ as the alloy coarsens.  In fact, we will present results that
suggest that the total elastic energy, $W_C+H$ , can actually decrease as the system coarsens. Importantly, this
means that elastic interactions that arise from atomistic misfit can destabilize an alloy; in other words,
they will enhance segregation.

We are not the first to consider some of the issues discussed here. For example, Eshelby \cite{ESH}
argued that for materials whose internal energy is elastic in origin the disordered state is unstable.
On the other hand, from Cahn \cite{CAHN} it would seem that the disordered state is stable.
De Fontaine \cite{DEF}  calls this the Elastic Energy Paradox.
 This paradox has also been observed by Cook and de Fontaine \cite{Cook1}
and Khachaturyan \cite{Khachaturyan} (see Chapter 13).
The physical origin of both the continuum elastic energy and the atomistic elastic enthalpy
is the same: different atomistic sizes of the alloy species. We shall see that
the source of confusion is in treating them separately.
They are both elastic energies, but they are operating on different length scales.

There is some experimental evidence that atomistic misfit will destabilize alloys. In the
case of thin film growth of a Si-Ge alloy on Si, Cullis et al \cite{CULLIS3} present
experimental evidence that lateral segregation occurs to lower strain.
In metallurgy, Hume-Rothery, Mabbott and Evans \cite{HMCE}, based on experimental observations,
proposed the ``15\% rule"  which states that binary solid solutions are very difficult to form
if the atomic size factor exceeds $\sim$ 15 \%. Eshelby \cite{ESH} asserted that this rule was a consequence of elastic instability
of the disordered state.
 In related work,
King \cite{KING} suggests that certain alloys based on copper are
unstable due to atomistic size effects.
Furthermore,
Woodilla and Averbach \cite{WA} report that the critical temperature for spinodal decomposition in experiments with Au-Ni
is $ \sim 220 ^\circ$, whereas Golding and Moss \cite{GM} used Cahn's approach and predicted it to be $\sim 0 ^\circ $ C.
Ren et al \cite{RWST} surmise that if
the enthalpy term was included it would raise this prediction to
more closely match experimental observations.

In this paper, we start with a well known atomistic model for a binary alloy in which the
bond energy is based on harmonic springs connecting atoms. The atoms are placed
on a simple square lattice with springs connecting nearest and
next to nearest neighbors with the equilibrium lengths being species dependent. First, we
derive the discrete equations for the displacement field and establish that if one
coarse-grains these equations one will recover the usual continuum equations of
elasticity.  This calculation reveals that continuum theory does
a good job at predicting average strain fields. Based on the form of
the discrete elastic energy one can propose a continuum version of the elastic
energy and  recover a well known and well used elastic energy which
we will denote $W_C$. Furthermore, we establish that the Euler-Lagrange equations
associated with $W_C$ yield the continuum equations that were derived
from coarse-graining the atomistic equations. In addition, if we
consider the case where continuum equations are isotropic, the elastic
energy of our alloy takes the same form as (\ref{eq:CAHN}).

However, a direct calculation of the elastic energy  of the atomistic model in mechanical equilibrium
reveals that its behavior can be quite different from its continuum counterpart and the key results  of this paper are the following:
\begin{enumerate}
\item This calculation demonstrates that the elastic energy is anisotropic and
scale dependent. Indeed, the calculation shows that in order to
evaluate the elastic energy one needs to understand the behavior of the
concentration field on all scales ranging from the atomistic to
the continuum.  The calculation shows that
expressions like Eq. (\ref{eq:CAHN}) can only be valid when the system is almost
completely segregated.

\item Furthermore,  it also shows that
the elastic energy can be changed by rearrangements at the atomistic scale
that would not affect the continuum concentration field.  
Segregation at the atomistic scale can lower the elastic energy and segregation is preferred over
intermixing if one accounts for the atomistic scale details. This is not captured by the continuum theory
which not only does not distinguish between the different configurations with microscopic segregation (due to lack of resolution)
but in fact predicts the opposite.

\item Our work demonstrates that it is critical to include the microscopic arrangements
in any elastic model to achieve even qualitatively correct behavior. Specifically, we show that the enthalpy  of mixing
$H$ depends both on the direction and wavenumber of the alloy's compositional variations.
 It is important to note that lack of microscopic information is a direct consequence
of the assumption that the stress free strain in the reference lattice is zero.
We demonstrate that even though the average stress free strain in the reference lattice is zero
the strain energy is not.

\end{enumerate} 

Interestingly, in the
modeling  of heteroepitaxial growth inclusion of any sort of elastic contribution
to the enthalpy of mixing is largely ignored (e.g. Spencer et al \cite{Spencer_2001} or Shenoy et al \cite{MHS}).
For models of spinodal decomposition (e.g. Cahn \cite{CAHN})
%$H$ is not usually explicitly included
 the enthalpy of mixing that includes the microscopic strain is not explicitly included
but it can be argued that it is implicitly included in the free energy term (see the discussion in the footnote on page
1478 of Ref. \cite{FPL}).
%In this case, it is assumed that $H=H(\theta)$,
 Even when enthalpy of mixing is included it is assumed to be a function of the macroscopic
concentration field (i.e, $H=H(\theta)$).
However, in view
of our discrete calculation it follows that $H$ must depend on the atomistic details of the alloy
and among other things must be scale dependent. 

Our calculations below are based on a ball and spring model but we surmise that they
are valid for real materials as well. To provide justification of this assertion
we also present calculations using density-functional theory applied to periodic
Si/Ge.  For example, we consider three dimensional checker-board patterns
with cubes of Si alternated with cubes of Ge
in which the size of the cubes is varied. The calculations show that as the sizes of the
cubes are increased the elastic energy is reduced - in agreement with the ball and
spring model. Other arrangements of alloys were also  considered and these also
make it clear that the elastic energy will depend on the atomistic
details of the alloy.

\section{Atomistic Alloy Model}
\begin{figure}[!ht]
\begin{center}
\includegraphics[width=3.0in]{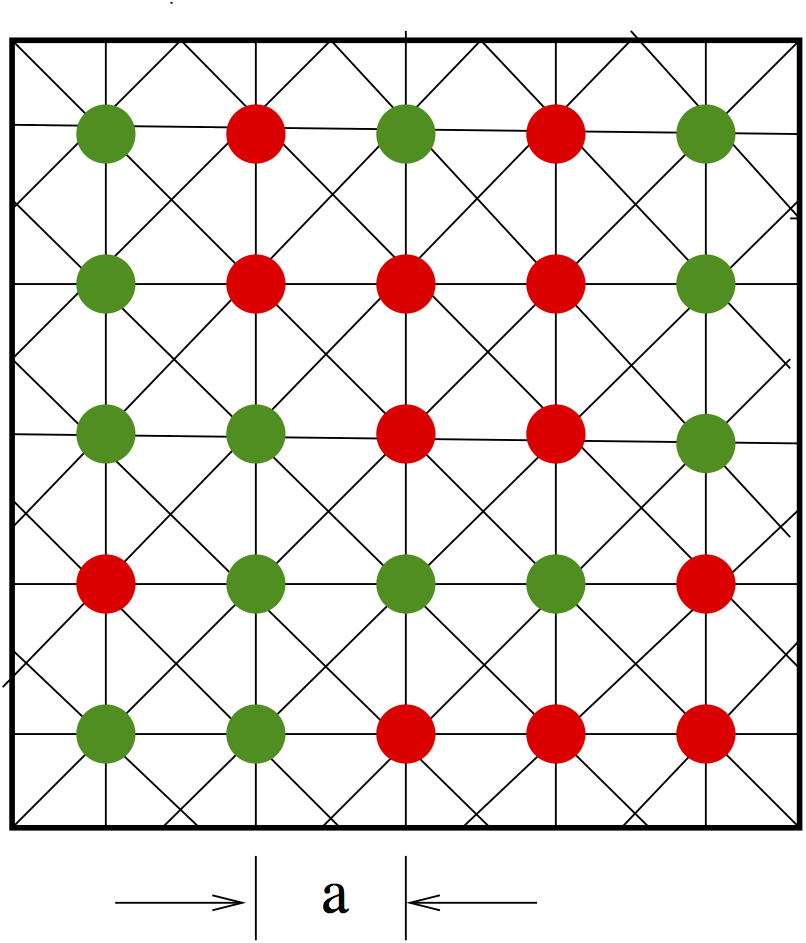}
\end{center}
\caption{(Color Online) The schematic of the ball and spring model when the atom are located on
the reference configuration}
\label{ALLOY}
\end{figure}

We consider a binary alloy with lattice mismatch (say Si/Ge for ease of exposition)
and use a ball spring model on a simple square lattice
in two dimensions with lattice spacing $a$ with periodic boundary conditions.
 The periodic domain is assumed to be a square domain of $N \times N$ lattice sites.
The atoms are connected by
Hookean springs between nearest and next-nearest neighbors with spring constants $K_L$ and $K_D$, respectively.
 For simplicity and easy of calculation the spring constants $K_L$ and $K_D$ are chosen to be independent of the type of bond
(Si-Si, Ge-Ge or Ge-Si). However for the Si/Ge system this is a reasonable assumption
for the elastic constants for the two species \cite{Wortman}.
The displacement of an atom at site $(\ell,j)$ from the reference
configuration is denoted $(u_{\ell,j},v_{\ell,j})$.
Each site on the lattice
will be occupied by either a Si or Ge atom (see Fig. \ref{ALLOY})
and we use the following indicator function to denote the atom type at site $(\ell,j)$,
\[
\theta_{\ell,j} = \left\{\begin{array}{cl}
                1 & {\rm if} \> (\ell,j) \> \mbox{if site contains a  Ge atom} \\
                0 & {\rm if} \> (\ell,j) \> \mbox{if site contains a  Si atom.}
             \end{array}
             \right. \nonumber\\
\]
The equilibrium
Si-Si and Ge-Ge bond lengths are denoted as $a_s$ and $a_g$,
whereas the bond length between a Si and a Ge atom  is taken to be
$\frac12(a_s+a_g)$. In this way the relative unstrained bond length between
atoms at sites $(\ell+n,j+m)$  and  $ (\ell,j)$ is given by
\begin{equation}
f_{\ell+n,j+m}=a_s-a+a_s\misfit(\theta_{\ell,j}+\theta_{\ell+n,j+m})/2,\label{BOND_LENGTH}
\end{equation}
where
\begin{equation}
\misfit = (a_g-a_s)/a_s.\label{misfit}
\end{equation}

The total elastic energy can then be written as
\begin{equation}
W=W(u,v,\theta)=\frac12\sum_{\ell=0}^{N-1}\sum_{j=0}^{N-1}
(w^{L}_{\ell,j}+w^{D}_{\ell,j}).\label{WEXACT}
\end{equation}
In this expression, the summand is the total elastic energy in springs connected to
the atom at site $(\ell,j)$ where
\begin{eqnarray*}
w^{L}_{\ell j}
 &=& \frac{K_L}{2}\sum_{n \in \{1,-1\} }
     ( \delta u_{\ell+n,j})^2+ (\delta v_{\ell,j+n})^2\quad\mbox{and}\\
w^{D}_{\ell j}
 &=&\frac{K_D}{4} \sum_{n,m \in \{-1,1 \}}
      (n\delta u_{\ell+n,j+m} + m\delta v_{\ell+n,j+m})^2,
\end{eqnarray*}
and
\begin{eqnarray}
\delta u_{\ell+n,j+m}= u_{\ell+n,j+m} - u_{\ell,j} - f_{\ell+n,j+m}\nonumber\\
\delta v_{\ell+n,j+m}= v_{\ell+n,j+m} - v_{\ell,j} - f_{\ell+n,j+m}.\nonumber
\end{eqnarray}

The model we choose is simple enough to be tractable but with the ability to incorporate the discrete
nature of the alloy, and to account for the microscopic arrangement of the atoms in an alloy.
In most models it is typically assumed that the alloy remains in mechanical equilibrium as
it evolves. This means that we need to evaluate the elastic energy, $W$, when the system is in mechanical equilibrium.
The equilibrium displacement field satisfies
\[
\frac{\partial W}{\partial u_{\ell,j}} =0 \quad\xtext{and}\quad
\frac{\partial W}{\partial v_{\ell,j}} =0.
\]
In other words
\begin{eqnarray}
&\phantom{+}&2 K_L (u_{\ell+1,j} - 2 u_{\ell,j } + u_{\ell-1,j }) \nonumber\\
               &+& K_D (u_{\ell+1, j+1} + u_{\ell-1, j+1 }
                     + u_{\ell+1,j-1 } + u_{\ell-1,j-1 }-4u_{\ell, j }) \nonumber\\
               &+&  K_D (v_{\ell+1,j+1}+v_{\ell-1,j-1}
                    - v_{\ell+1, j-1}-v_{\ell-1, j+1})\nonumber\\
&=&\misfit a_s\left[ K_L(\theta_{\ell+1,j} - \theta_{\ell-1,j})+K_D(
\theta_{\ell+1,j+1}
+\theta_{\ell+1,j-1}
-\theta_{\ell-1,j+1}
-\theta_{\ell-1,j-1})\right]\label{FX}
\end{eqnarray}
and
\begin{eqnarray}
   &\phantom{+}& 2K_L (v_{\ell, j+1 } - 2 v_{\ell, j } + v_{\ell, j-1 }) \nonumber \\
               & +& K_D(v_{\ell+1,j+1 } + v_{\ell-1,j+1 }
            + v_{\ell +1, j-1} + v_{\ell-1 ,j-1}- 4v_{\ell, j}) \nonumber \\
              & +&  K_D (u_{\ell+1, j+1 } + u_{\ell-1, j-1}
                  - u_{\ell+1, j-1 } - u_{\ell-1, j+1}) \nonumber\\
&=&\misfit a_s\left[ K_L(\theta_{\ell,j+1} - \theta_{\ell,j-1})+K_D(
\theta_{\ell+1,j+1}
+\theta_{\ell-1,j+1}
-\theta_{\ell+1,j-1}
-\theta_{\ell-1,j-1})\right].\label{FY}
\end{eqnarray}
 It is easy to see from Eq. (\ref{WEXACT}) that $ \displaystyle \frac{\partial^2 W}{\partial u_{\ell,j}^2},\frac{\partial^2 W}{\partial v_{\ell,j}^2} > 0$
and $W \to \infty$ as $ u_{\ell,j},v_{\ell,j} \to \pm \infty$ hence the solution to Eq. (\ref{FX}) and Eq. (\ref{FY}) is the unique minimizer of the $W$.
Later in this paper we shall evaluate $W$ when the displacement field satisfies Eqs. (\ref{FX}) and (\ref{FY}).
However, now it is useful to look at the continuum limit of these equations.

\section{Continuum Limit}

For alloys, $\theta$ can vary on the scale of the lattice and cannot be
used as a continuum variable. Instead we appeal to a coarse-grained value:
\[
\bar{\theta}_{\ell,j} =  \sum_{\ell',j'}A_{R}(\ell-\ell',j-j')\theta_{\ell',j'} \quad \mbox{with} \quad
\sum_{\ell',j'}A_{R} = 1
\]
where $A_R$ is an averaging kernel and $R$ is a length scale over which the averaging takes place
(e.g. $A_R(\ell,j)= C \exp[-(\ell^2+j^2)/R^2 ] $ where $C^{-1}= \sum_{\ell,j} \exp[-(\ell^2+j^2)/R^2 ]). $
Since the coarse-grained variables are smooth functions of
of the lattice site, we can introduce a smooth function $\Theta$ such that
$\bar{\theta}_{\ell,j} = \Theta(a\ell,aj)$.

\subsection{The Displacement Field}

We apply the coarse-graining operation to the equations governing the displacement field
(i.e.  Eqs. (\ref{FX}) and (\ref{FY})). Since  the equations of  mechanical equilibrium are linear,
 it is clear that the coarse-grained variables, $(\bar{u}_{\ell,j},\bar{v}_{\ell,j})$
satisfy the same equations as the atomistic system. However, unlike their atomistic counterparts the
coarse grained variables  vary slowly over atomistic scales such that $\bar{u}_{\ell,j} =U(a\ell,a j)$
and $\bar{v}_{\ell,j} =V(a\ell,a j)$,  where $U$ and $V$ are smooth functions.
We can consequently make use of approximations such as
\[
\bar{u}_{\ell+1,j} - 2 \bar{u}_{\ell,j } + \bar{u}_{\ell-1,j } \approx a^2 U_{xx}
\]
to find that continuum variables satisfy the following equations

\begin{eqnarray}
K_L {U}_{xx} + K_D ( {U}_{xx} + U_{yy} + 2{V}_{xy})  &=& (a_g-a_s)(2K_D+K_L)\Theta_x,\nonumber\\[-.1in]
\phantom{o}\hspace{-1in} \hspace{2in}&&\label{CON_ELAS} \\[-.1in]
K_L{V}_{yy} + K_D ( V_{xx} + {V}_{yy} +  2U_{xy})  &=& (a_g-a_s)(2 K_D+K_L)\Theta_y.\nonumber
\end{eqnarray}

Therefore, continuum theory can predict the coarse-grained displacement field
in terms of the coarse-grained concentration field. Finally, we point out that
for the case $K_L=2K_D$ the system (\ref{CON_ELAS}) corresponds to isotropic elasticity.
 This is seen by noting that the above equations represent the constitutive relation between the stress
and the strain and collecting the elasticity tensor (see \cite{Schulze_Smereka,FPL}). 

\subsection{Continuum Elastic Energy}

The elastic energy is a quadratic function of the displacement field and accordingly
it is not a simple matter to apply the coarse-graining operation to $W$ and express
the result in terms of the coarse grained variables. Instead, what can be done is
to use approximations such as
\begin{equation}
\theta_{\ell+1,j} - \theta_{\ell,j } \approx a \Theta_{x}\label{DER_APP}
\end{equation}
and replace the atomistic value of $\theta$ by its continuum value, $\Theta$.
Given that the atomistic values are not necessarily smooth functions of the
lattice site, approximations of the type given by (\ref{DER_APP}) could be rather poor.

Nevertheless, if we apply this procedure to the atomistic energy (Eq. \ref{WEXACT}),
we arrive at the following continuum version of the elastic energy :

\begin{equation}
W_C=\frac12 \int \left[\, (K_D+K_L)(S_{xx}^2+ S_{yy}^2) + 2 K_D S_{xx} S_{yy} +
4 K_D S_{xy}^2\,\right] \,dxdy,\label{WWWC}
\end{equation}

where $S = E - E^0$,
\[
E_{jk}= \frac12\left(\frac{\partial U_j}{\partial x_k}+
\frac{\partial U_k}{\partial x_j}\right)
\]
and
\[
E^0= \frac{1}{a}\left(\begin{array}{ll}  a_s-a + (a_{g}-a_{s})\Theta   & 0\\
0 & a_s-a + (a_{g}-a_{s})\Theta \end{array}\right).
\]
We remark that $E^0$ is sometimes called the stress free strain. It should be
emphasized that Eq. (\ref{WWWC}) is a commonly used model (e.g. Refs. \cite{Spencer_2001, Tu_2004,MHS}).
In addition, the
Euler-Lagrange equations for $W_C$ give rise to
the equations for continuum elasticity (\ref{CON_ELAS}). By this we mean
\[
\delta W_C = 0\quad \Rightarrow\quad \sum_{k=1}^2 \frac{\partial}{\partial x_k} \frac{\partial W_C}{\partial U_{j,k} } =0
\]
will yield (\ref{CON_ELAS}).

By following the same approach as Cahn \cite{CAHN}, we will compute the
elastic energy of an alloy in mechanical equilibrium for the isotropic case ($K_L=2K_D$). We consider
a periodic region of size $2\pi\times 2\pi $ and we find
\[
W_C= \overline{W} + \widetilde{W}_C,
\]
where
\[
\overline{W}= 4 K_D (a_s-a+ \Theta_0(a_g-a_s))^2 (2\pi)^2/a^2,
\]
and
\[
\widetilde{W}_C =
\frac43K_D \left(\frac{\mu a_s}{a}\right)^2 \int\int
(\Theta(x,y)-\Theta_0)^2\, dxdy
\]
where $\Theta_0$ = average value of $\Theta$. We note that
$\overline{W}$ is the ``DC'' contribution (Direct Current or non-oscillating)  to $W_C$
and $\widetilde{W}_C$ is the contribution from the compositional variations.
Notice that by adjusting the lattice spacing $a$ of the reference configuration
we can make $\overline{W} = 0$. This value is $a=a_s+\Theta_0(a_g-a_s)$ and is sometimes
called Vegard's Law. An important conclusion from this continuum formulation is
that the elastic energy is zero for a homogenous alloy whose reference
lattice spacing satisfies Vegard's Law.

\section{Elastic Energy in the Atomistic Case}

It should be pointed out that
here we are closely following the calculations of Cahn \cite{CAHN}.
In this section we apply his approach to discrete equations while he considered
continuum equations.

Since we are in the periodic
setting it is useful to  expand various quantities in
a discrete Fourier series. For example
%\[
%u_{\ell,j} =\frac{1}{N^2}\sum_{n=0}^{N-1} \sum_{m=0}^{N-1} \hat u_{m,n}e^{i \alpha (m\ell+nj )}
%\]
%\[
%v_{\ell,j} =\frac{1}{N^2}\sum_{n=0}^{N-1} \sum_{m=0}^{N-1} \hat v_{m,n}e^{i \alpha (m\ell+nj )}
%\]
\begin{equation}
\theta_{\ell,j} =\frac{1}{N^2}\sum_{n=0}^{N-1} \sum_{m=0}^{N-1} \hat \theta_{m,n}e^{i \alpha (m\ell+nj )}
\label{FS}
\end{equation}
where $\alpha =  2\pi/N$ and $\hat \theta_{m,n}$ is the
discrete Fourier transform of $\theta_{\ell,j}$.
Furthermore, we will for now restrict our calculations to the case $K_L=2K_D$, in which
case the continuum limit is isotropic.
Applying the discrete Fourier transform to (\ref{FX}) and (\ref{FY}) and solving for the
transformed displacement field we find
\begin{equation}
\hat u_{m,n} =\left\{\begin{array}{ll}  0 & n=m=0\\
\displaystyle{\frac{-2\misfit a_s i \hat\theta_{m,n}\cos^2(n\alpha/2)\sin(m\alpha)}{4-\cos(m\alpha)-\cos(n\alpha)-
2\cos(m\alpha)\cos(n\alpha) }} & \xtext{otherwise}
\end{array}\right.\label{UHAT}
\end{equation}
and
\begin{equation}
\hat v_{m,n} =\left\{\begin{array}{ll}  0 & n=m=0\\
\displaystyle{\frac{-2\misfit a_s i\hat\theta_{m,n}\cos^2(m\alpha/2)\sin(n\alpha)}
{4-\cos(m\alpha)-\cos(n\alpha)- 2\cos(m\alpha)\cos(n\alpha)}} & \xtext{otherwise,}
\end{array}\right.\label{VHAT}
\end{equation}
where $\hat u_{m,n}$ and $\hat v_{m,n}$ are the discrete Fourier transforms of
$u_{\ell,j}$ and $v_{\ell,j}$, respectively.

Our goal is now to calculate the total elastic energy of the ball and spring
model when the system is in mechanical equilibrium.  To that end, it is useful to
define the average value of $\theta$ :
\[ \theta_0 = \mathop{\sum_{n=0}^{N-1}\sum_{m=0}^{N-1}}_{(m,n)\ne(0,0)} \theta_{m,n}.\]
Now it is worth noting that $\theta_0 = \hat{\theta}_{0,0}$. 
 In view of the
two formulas above (\ref{UHAT}) and (\ref{VHAT}) it is apparent that each component of
the displacement field has mean zero. Therefore, in mechanical equilibrium
we have
\[
W(u,v,\theta)=W(0,0,\theta_0)+W(u,v,\theta-\theta_0)\equiv \overline{W}+\widetilde{W}
\]
where $\overline{W} =N^24 K_D(a_s-a+ \theta_0(a_g-a_s))^2$.
The first term, $\overline{W}$, represents the contribution to the elastic energy if the
material was of a uniform concentration. The second term, $\widetilde{W}$, results
from the compositional variations.
We compute $\widetilde{W}$ using
Parseval's formula combined with (\ref{UHAT}) and (\ref{VHAT}).
A lengthy calculation reveals
\begin{equation}
\widetilde{W}=
\frac{K_D (a_s \misfit)^2}{N^2}\mathop{\sum_{n=0}^{N-1}\sum_{m=0}^{N-1}}_{(m,n)\ne(0,0)}
 G(\alpha m,\alpha n)
|\hat{\theta}_{m,n}|^2 \label{EXACT}
\end{equation}
where
\[
G(x,y)=\frac{T(x,y)}{8(4-\cos x-\cos y- 2\cos x \cos y)^2}
\]
and
\begin{eqnarray}
T(x,y) &=& 196 -87 [\cos x +\cos y]+ 4 [\cos 2x +\cos 2y]+ \cos 3x +\cos 3y\nonumber\\
&-&42 [ \cos (x-y) + \cos (x+y) ] + 2 [ \cos(2x+2y)+\cos(2x-2y) ]\nonumber\\
&+&11 [ \cos (2x+y) +\cos(x+2y) + \cos(2x-y) +\cos(x-2y) ]\nonumber\\
&+&\cos(x-3y)+\cos(3x-y)+ \cos(x+3y) + \cos(3x+y) . \nonumber
\end{eqnarray}
Since  $G(2\pi-k_x,k_y)$ = $G(k_x, 2\pi-k_y)$ = $G(2\pi-k_x,2\pi- k_y)$=$G(k_x,k_y)$ it is enough
to consider $G(k_x,k_y)$ for $0 < k_x \le \pi$ and $0 < k_y \le \pi$.

Notice that $ \widetilde{W}$ does not depend on $a$ but $\overline W$ does. In fact
we can relax our alloy by changing $a$: If we pick $a=a_s(1+\theta_0 \mu)$ (Vegard's law), then
$\overline W=0$. For the remainder this value of $a$ will be used. Therefore
the elastic energy of the relaxed alloy is $\widetilde{W}$.

\begin{figure}[!ht]
\begin{center}
\includegraphics[width=3.0in]{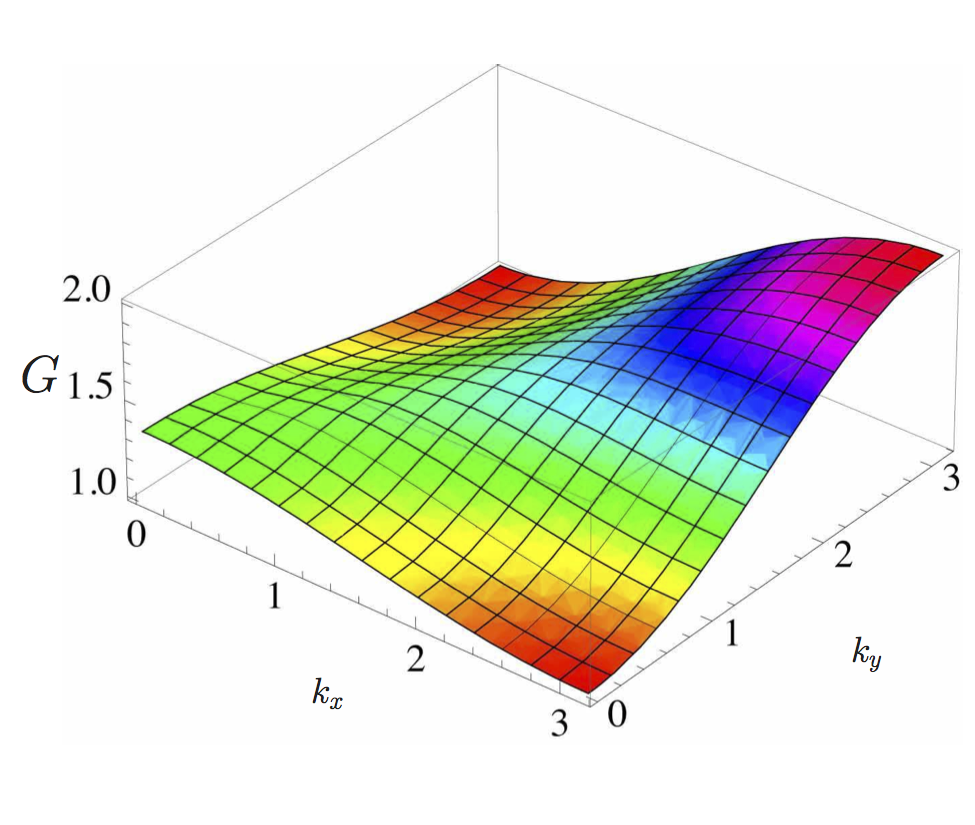}
\end{center}
\caption{(Color Online) A plot of $G(k_x,k_y)$  for the case $K_L=2 K_D\, (K_D=1)$ }
\label{GGG}
\end{figure}

The dependence of
$\widetilde{W}$ on $\theta$ can be gleaned from Fig.~\ref{GGG}.
Note that $\widetilde{W}$ is a weighted sum of the $G(k_x,k_y)$ with weights $|\hat\theta|^2$.
Different regions of the $k$-space represent different aspects of composition profiles.
For example, an alloy that is finely intermixed on an atomic scale has
weights $|\hat\theta|^2$ concentrated near $(k_x,k_y)=(\pi,\pi)$, while for an alloy with variations on continuum length scales
the weights are concentrated near $(k_x,k_y)=(0,0)$. It therefore follows from Fig.~\ref{GGG} that finely mixed alloys have more
elastic energy than those that are segregated.
Moving along from  $(\pi,\pi)$ to $(0,0)$ corresponds to a coarsening patterns with checker board symmetry.
The fact that $G(\pi,\pi)/G(0,0)= 3/2$ indicates that intermixing can significantly
increase the elastic energy.
Interestingly, it also follows that fine line patterns (along $k_x =0$ or $k_y=0$ axis) have even less elastic energy for the  case where
the constants $K_L$ and $K_D$ are related by $K_L=2K_D$.
Although these cases can be viewed as different regions of $k$-space, typical composition profiles would of course have weights in all regions and
no separation of scales. Thus, Fig.~\ref{GGG} also points to the difficulty relying on formulas like Eq. (\ref{MME}), since it is clear
that the elastic energy depends on the details of the atomistic arrangement over the full range of length scales.

As mentioned above, the behavior near $(k_x,k_y)=(0,0)$ describes an alloy with variations on continuum length scales
and we note that
\begin{equation}
G(\alpha x, \alpha y)= \frac43 + \frac{3x^4-10x^2y^2+3y^4}{9(x^2+y^2)}\alpha^2 + O(\alpha^4) , \label{CONTLIM}
\end{equation}
which means the limit $\lim_{(k_x,k_y) \to (0,0)} G(k_x,k_y)$ is  well defined and equal to $\frac 43$.
This is consistent with the assertion that the case $K_L=2K_D$  recovers isotropic elasticity.
Furthermore if it assumed that $\theta_{\ell,j}$ is a continuum variable (i.e. $|\hat \theta_{m,n}|$ is strongly concentrated
at the origin), then
\begin{eqnarray}
\lim_{\alpha\to 0} \widetilde{W} &=&
\frac{4 k_D (a_s \misfit)^2}{3 N^2}\mathop{\sum_{n=0}^{N-1} \sum_{m=0}^{N-1}}_{(m,n)\ne (0,0)}
|\hat{\theta}_{m,n}|^2\nonumber\\ &=&
\frac43k_D (a_s \misfit)^2\sum_{\ell=0}^{N-1} \sum_{j=0}^{N-1}
(\theta_{\ell,j}-\theta_0)^2 . \label{CONTINUUM}
\end{eqnarray}
%One can also infer that $\lim_{k \to 0} G(k) =\frac43$, which means that
%for $K_L=2K_D$ the ball and spring model is isotropic in the continuum limit,
%where the length scales of the composition changes are much larger than the atomic scale.
Therefore, the model is anisotropic on small scales and
isotropic on large scales.  The anisotropy of ball and spring models on atomistic scales
has been discussed in Refs.~\cite{Cook1, FP}.

\subsection{Anisotropic Continuum Limit}
\begin{figure}[!ht]
\begin{center}
\includegraphics[width=3.0in]{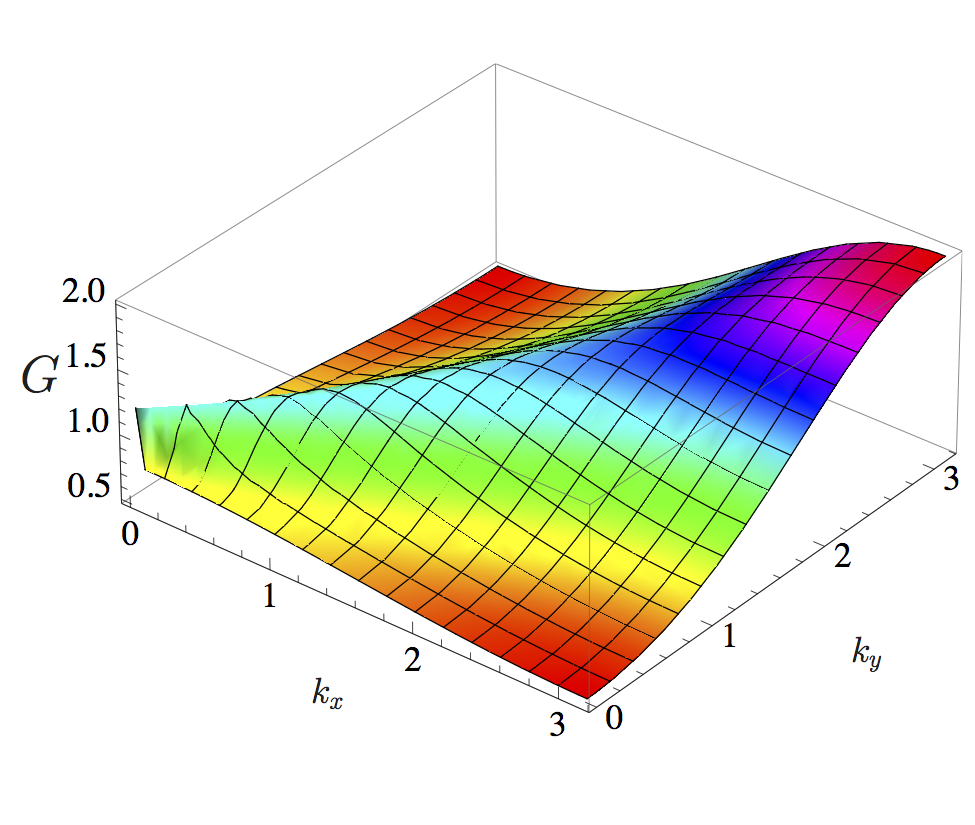}
\end{center}
\caption{(Color Online) A plot of $G(k_x,k_y)$ for $K_L=K_D=1$}
\label{GGGKK}
\end{figure}

\begin{figure}[!ht]
\begin{center}
\includegraphics[width=3.0in]{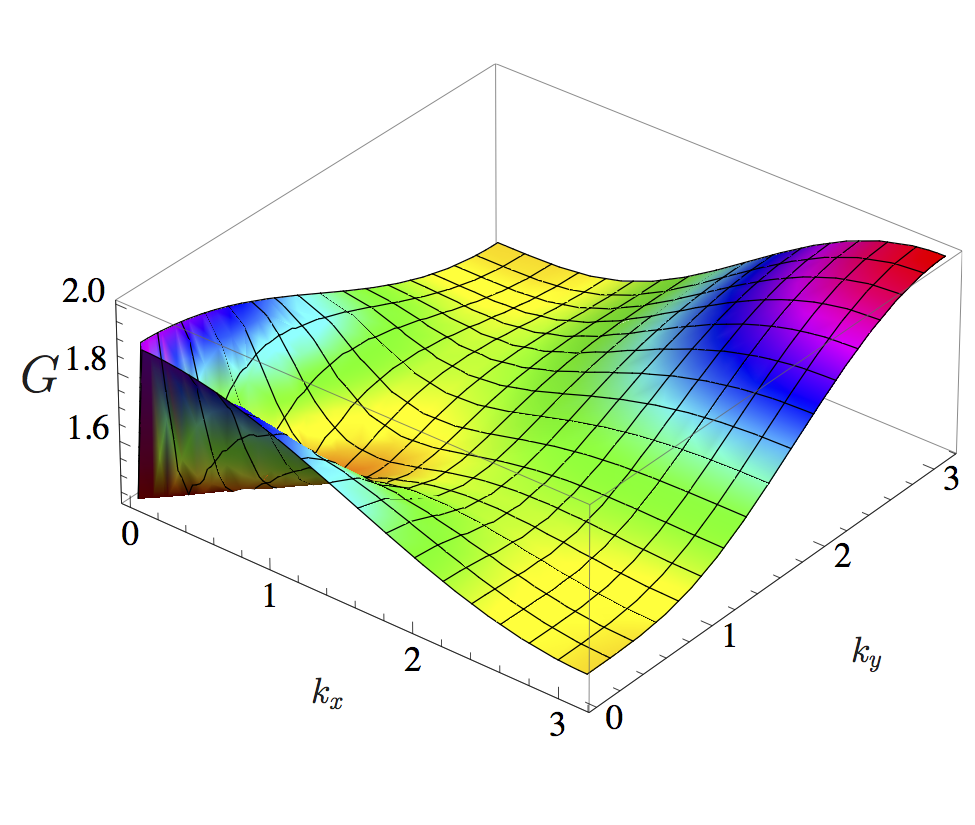}
\end{center}
\caption{(Color Online) A plot of $G(k_x,k_y)$ for $K_L=3 K_D\, (K_D=1)$}
\label{GGGK3K}
\end{figure}

We note that for $K_L \ne 2K_D$ the continuum limit becomes
anisotropic.  It is well known (see \cite{FPL} page 1447) that in this case the function
$G(k_x,k_y)$ is a homogeneous function of order 0 near the origin and
$\lim_{(k_x,k_y) \to (0,0)} G(k_x,k_y)$ is not well defined giving
rise to the cusp-like behavior near the origin (see Figures \ref{GGGKK} and \ref{GGGK3K}).

\begin{figure}[!ht]
\begin{center}
\includegraphics[width=6.0in]{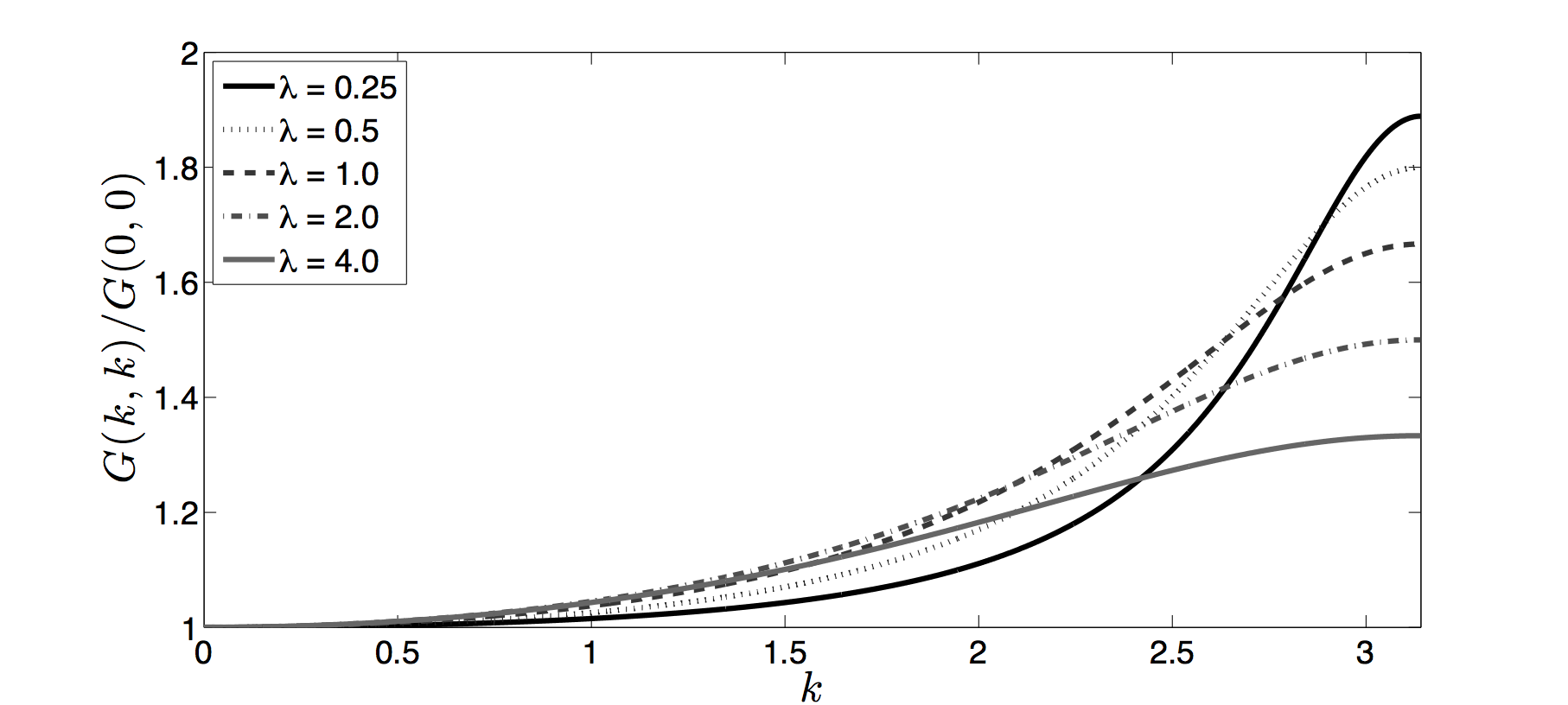}
\end{center}
\caption{(Color Online) A plot of $G(k,k)/G(0,0)$ vs $k$ for different values of $\lambda = K_L/K_D$}
\label{GKK}
\end{figure}

The nature of $G$
makes it difficult to make claims
about the relative amount of
elastic energy in the different scales
for a wide range of values of $K_L$ and $K_D$.
However, the behavior along the diagonal is
relatively robust, as can be seen in Figure \ref{GKK}.
In addition, we can establish that
for arbitrary $K_L$ and $K_D$ one has
$\lim_{\alpha\to 0} G(\pi,\pi)/G(\alpha,\alpha)= (K_L+4K_D)/(K_L+2K_D) > 1$,
which suggests that the tendency for a finely mixed systems
to have a greater elastic energy than a segregated one persists.

\subsection{Elastic Enthalpy of Mixing}

If we compare Eqs. (\ref{EXACT}) and (\ref{CONTINUUM}), it is evident that if one
ignores the microscopic scales of the concentration field (i.e. by only considering the
Fourier modes near the origin) then the discrete elastic energy is consistent with the
continuum elastic energy.  But more importantly the discrete and continuum cases are different
because the continuum elastic energy fails to account for the contribution of the microscopic
scales.  The reason the coarse grained displacement field does not suffer from this fate is that
the microscopic strain fields average out. Since the energy is a quadratic quantity, the
microscopic contributions do not cancel when averaged.

As can be inferred by Tsao \cite{TSAO}, atomistic variations of the alloy concentration
lead to microscopic strain whose contribution to the elastic energy can be referred to
as the elastic contribution to the enthalpy of mixing, $H$. Therefore, it follows
from Eq. (\ref{EXACT}) that $H$
is determined by both the direction and wavelength of the Fourier modes
of the microscopic concentration field. Furthermore, it appears that $H$ cannot be a function
of the average concentration unless some simplifying approximations are made. This can be done
in one setting, namely if one assumes that the alloy is in local thermodynamic equilibrium. In this
case one has, in principle, $H=H(\Theta,T)$. However, when the alloy is not in local thermodynamic equilibrium,
as is the case in epitaxial growth or spinodal decomposition, then
$H \ne H(\Theta,T)$, and then $H$ will be determined by details at the atomistic scale. Unfortunately
this information is completely lost during coarse graining. This points to the difficulty of
modeling elastic energy using continuum theory.

%\begin{figure}
%\begin{picture}(400,300)
%\put(0,130){\psfig{figure=checker.png,width=1.7in}}
%\put(0,0){\psfig{figure=checker2.png,width=1.7in}}
%\put(140, 120){ $\Rightarrow$}
%\put(160, 120){ Continuum limit }
%\put(255, 120){ $\Rightarrow$}
%\put(280,70){\psfig{figure=continuum.png,width=1.7in}}
%\end{picture}
%\caption{The figure on the left shows two microscopic arrangements that yield the same
%continuum alloy concentration (shown on the right). The upper left hand figure shows a $1\times 1$ checkerboard
%pattern whereas the lower one is a $2 \times 2$ pattern.}
%\label{CHECKER}
%\end{figure}
%\begin{figure}
%\begin{picture}(400,300)
%\put(0,130){\psfig{figure=checker.png,width=1.7in}}
%\put(0,0){\psfig{figure=checker2.png,width=1.7in}}
%\put(140, 120){ $\Rightarrow$}
%\put(160, 120){ Continuum limit }
%\put(255, 120){ $\Rightarrow$}
%\put(280,70){\psfig{figure=continuum.png,width=1.7in}}
%\end{picture}
%\caption{The figure on the left shows two microscopic arrangements that yield the same
%continuum alloy concentration (shown on the right). The upper left hand figure shows a $1\times 1$ checkerboard
%pattern whereas the lower one is a $2 \times 2$ pattern.}
%\label{CHECKER}
%\end{figure}

\begin{figure}[!ht]
\begin{center}
\includegraphics[width=5.6in]{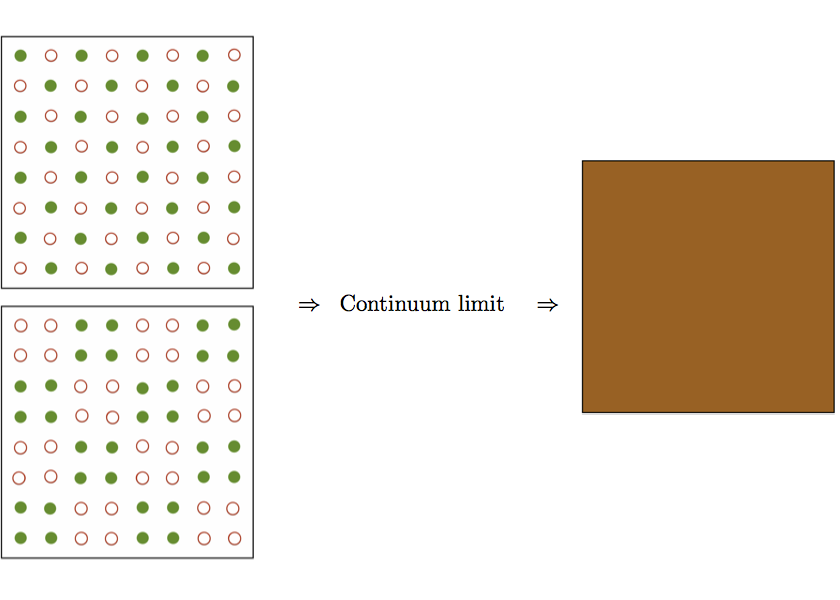}
\end{center}
\caption{(Color Online) The figure on the left shows two microscopic arrangements that yield the same
continuum alloy concentration (shown on the right). The upper left hand figure shows a $1\times 1$ checkerboard
pattern whereas the lower one is a $2 \times 2$ pattern.}
\label{CHECKER}

\end{figure}

The following example is useful in elucidating some of the issues discussed above. Consider a binary alloy
arranged on a checker board pattern as shown in Fig. \ref{CHECKER} ($\theta_{\ell,j}=\frac12(1-(-1)^{\ell+j})$).
We have chosen the lattice spacing of
the reference configuration to be in accordance with Vegard's Law ($a=\frac12(a_s+a_g)$). Clearly then on a continuum
scale the alloy can be considered homogeneous $\Theta(x,y) = \Theta_0$ with $\Theta_0=\frac12$. Consequently,
the widely used continuum formulation (Eq. \ref{eq:CAHN}) predicts that the elastic energy is zero.
On the other hand, it is clear that there will be elastic energy in the bonds due to fact that the atoms on this lattice
do not correspond to their equilibrium bond length.
The continuum approximation ignores the microscopic stress fields.
This is fine for the coarse-grained displacement field as the microscopic
fields average out.  However, when computing the energy the microscopic fields do not average out.

Now one could of course assert that the  energy associated with the microscopic strain, calculated above as $\widetilde{W}$ in Eq. (\ref{EXACT}) is the enthalpic component
 $H$
of the elastic energy and in many respects it is. However, we will now argue that it cannot
be simply characterized by the continuum value of the concentration.
This can be seen as follows: Suppose we now coarsen the checker board
(see Fig. \ref{CHECKER}) so that the length scale is 2 atomistic units. Clearly on the continuum scale
the alloy can still be considered homogeneous and (Eq. \ref{eq:CAHN}) still predicts
that the elastic energy is zero. However, if we appeal to Fig. \ref{GGG} we can infer that
the elastic energy of the $2\times 2$ checker board will be smaller than that of the $1 \times 1$ checker board,
but the continuum value of the alloy concentration has not changed. Therefore, we
conclude that the elastic mixing enthalpy of a binary alloy cannot be characterized in
terms of the alloy concentration alone. Of course Fig. \ref{GGG} makes this quite clear.

\section{Density Functional Theory}

%\begin{figure}
%\begin{picture}(300,120)
%\put(0,0){\psfig{figure=dft1.png,width=1.7in}}
%\put(180,0){\psfig{figure=dft2.png,width=1.7in}}
%\end{picture}
%\caption{The atom arrangements for the density functional calculations.
%The figure on the right shows the $N_C=1$ checkerboard pattern whereas the one on the
%left shows the $N_C=64$ pattern}
%\end{figure}

\begin{figure}
\includegraphics[width=3.5in]{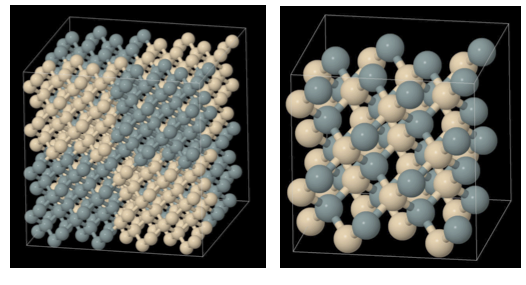}
\caption{(Color Online) The atom arrangements for the density functional calculations.
The figure on the right shows the $N_C=1$ checkerboard pattern whereas the one on the
left shows the $N_C=64$ pattern}
\end{figure}

To explore the broader applicability of the ideas discussed above, we perform DFT calculations for Ge/Si (with a zinc-blende lattice).
%as well as Ag/Pt (with an fcc lattice).
All results presented here were obtained with the FHI-AIMS code \cite{DFT1}.
This is an all-electron full potential DFT code that uses numeric atom centered orbitals as its basis set. We have carefully tested convergence of our results with respect to the basis set, and the density of the (numerical) integration mesh, and have used parameters as they are implemented in FHI-AIMS in the default setting ``tight'' \cite{DFT1}. We use GGA-PBE for the exchange-correlation functional \cite{DFT2}. All calculations were done with supercells with periodic boundary conditions, and the geometric configurations are fully relaxed.

Before we discuss the DFT results we note that one cannot simply repeat the calculation done in the case of the ball and spring model.
The reason is that DFT provides one with the total energy for a given configuration, which has no clear separation between bond and elastic energy. However, for Si or Ge bulk and Si-Ge in a perfect unstrained lattice the only contribution to the energy is the chemical bond energy.
If we assume that each atom forms exactly 4 bonds with its nearest neighbors (which is reasonable for a zinc-blende structure),
we can estimate the Si-Si, Ge-Ge, and Si-Ge bond-strengths $E^B_{\rm Si}$, $E^B_{\rm Ge}$, and $E^B_{\rm SiGe}$ as follows:
We calculate the energy of a bulk system with $N_{total}$ atoms, subtract the energy of $N_{total}$ isolated atoms, and divide this number by $2N_{total}$. In a strained system, the elastic contribution $E^{\rm el}$ is defined as the difference between the total energy of the strained system and the sum of all nearest neighbor bonds.
This definition of $E^{\rm el}$ is consistent with assuming that during the mechanical relaxation
the bond energy remains the same while the elastic energy changes and that the total energy
is the sum of elastic and bond energies. All of our calculations are at $T=0$, so there are no entropic contributions.

\begin{table}[ht]
  \caption{ Bond strengths $E^B$ (in eV) calculated with DFT.}
\begin{center}
 \begin{tabular}{| l | c | c | c | c| }
 \hline
  lattice constant & $E^B_{\rm Si}$ & $E^B_{\rm Ge}$ & $E^B_{\rm SiGe}$ & $\frac12 (E^B_{\rm Si} + E^B_{\rm Ge} )$ \\
  \hline
  a$_{\rm Si}$ & 2.677 & 2.176 & 2.431 & 2.426 \\
  a$_{\rm SiGe}$ & 2.661 & 2.213 & 2.442 & 2.437 \\
  a$_{\rm Ge}$ & 2.608 & 2.227 & 2.420 & 2.418 \\
 \hline
\end{tabular}
\label{table:bond_strength}
\end{center}
\end{table}
We calculated the bond energies for systems where all atoms are in a lattice with the optimized lattice constants a$_{\rm Si}$, a$_{\rm Ge}$, and $a_{\rm SiGe}$. 
 The optimized lattice constants are calculated by optimizing three different configurations. The first consists purely of Si atoms. The second purely of Ge atoms.
The third system used to calculate $a_{\rm SiGe}$ consists of 50\% Si and 50 \% Ge arranged in an alternating fashion so that all bonds in the zinc blend structure are
Si-Ge bonds and are configurationally equivalent to one another under periodic boundary conditions. Thus the uniform lattice spacing of the optimized structure yields the $a_{\rm SiGe}$ lattice spacing. 
The three different bond strengths are calculated in a similar manner. 
The results are shown in table~\ref{table:bond_strength}.
For all lattice constants we find that $E^B_{\rm Si}$ is stronger than $E^B_{\rm Ge}$, while $E^B_{\rm SiGe}$ is in-between. But a closer inspection of the numbers reveals that $E^B_{\rm SiGe}$ is 2 to 5 meV stronger than the average of $E^B_{\rm Si}$ and $E^B_{\rm Ge}$ for all lattice constants. This implies that simple bond counting arguments predict that a perfectly intermixed Si/Ge system is always preferred, regardless of whether the system has the lattice constant a$_{\rm SiGe}$, is compressed to a$_{\rm Si}$, or is stretched to a$_{\rm Ge}$.

Now we proceed to understand the system with elastic effects.
For this we consider various configurations of a $\rm Si_{0.5}Ge_{0.5}$ alloy.
We compare bulk alloy composition profiles that resemble a 3-dimensional checkerboard, where each ``checker-unit'' consists of 1, 8, or 64 atoms of the same type (i.e., all Si or all Ge)
occupying a cubic region in space (see Fig. \ref{GGG}). This corresponds to moving along the diagonal for $G$ (cf. Fig. \ref{GGG}). We do this by placing the configuration in a zinc blend lattice with reference lattice constant a$_{\rm Si}$ or a$_{\rm SiGe}$ and then optimizing the structure.

Table~\ref{table:checker} summarizes the DFT results.  $\Delta E^{\rm tot}$ is the difference in the total energy ( $E^{\rm tot}$ that includes bond and elastic energies) for systems with checker units that consist of $N_C$ atoms, and   a checkerboard with $N_C = 1$. For a system with the lattice constant a$_{\rm Si}$ (which is most relevant for Ge deposition on Si) we find that the checkerboard with $N_C = 8$ is preferred by 1 meV per atom (over a system with $N_C =1$), and that one with $N_C = 64$ is preferred by 4 meV per atom. For the lattice constant a$_{\rm SiGe}$, the coarser system with $N_C = 64$ is also preferred, but only by 1 meV.

The numbers for $\Delta E^{\rm el}(N_C) = E^{\rm el}(N_C) - E^{\rm el}(N_C=1)$ represent the change in energy per atom after correcting for the fact that simple bond counting arguments favor intermixing and should be considered the true elastic contribution due to coarsening. For example, for a system with $N_C =8$, half of the bonds are converted from being Si-Ge bonds to being either Si-Si or Ge-Ge bonds, and one can show that $\Delta E^{\rm el} = \Delta E^{\rm tot} - 1/2 E^B_{\rm Si} - 1/2 E^B_{\rm Ge} + E^B_{\rm SiGe}$.
Since the average of the Si-Si and Ge-Ge bond is 5 meV weaker than the Si-Ge bond (cf. table~\ref{table:bond_strength}), intermixing is preferred by 5 meV per atom for this system. Similarly, for a system with $N_C = 64$, $\Delta E^{\rm el} = \Delta E^{\rm tot} - 2/3 E^B_{\rm Si} - 2/3 E^B_{\rm Ge} + 4/3 E^B_{\rm SiGe}$, and intermixing is preferred by 7 meV per atom. $\Delta E^{\rm el}$ then represents the fact that elastic contributions have to overcome this favoring of intermixing. The results confirm that elastic effects favor segregation.

 We further note that the numbers given in table~\ref{table:bond_strength} and~\ref{table:checker} are real numbers. The units are in eV.
The accuracy of the DFT calculations is at best of the order of meV.
Therefore, we have rounded the results to 3 significant digits past
the decimal point (i.e., meV). In table~\ref{table:checker} we then report differences,
and these are given in meV.

 \begin{table}[ht]
  \caption{ $\Delta E^{\rm tot}$ and $\Delta E^{\rm el}$ for different values of $N_C$. Energy changes (in meV) are with respect to a system with $N_C=1$ (a perfect zinc-blende structure).}
  \begin{center}
 \begin{tabular}{| l | c | c |c|c| }
 \hline
  $N_C$ & $\Delta E^{\rm tot}(a_{\rm Si})$ & $\Delta E^{\rm tot}(a_{\rm SiGe})$
 & $\Delta E^{\rm el}(a_{\rm Si}$) & $\Delta E^{\rm el}(a_{\rm SiGe})$ \\
  \hline
  1 & 0 & 0 & &  \\
  8 & -1 & 1 & -6 & -4 \\
  64 & -4 & -1 & -11 & -8 \\
 \hline
\end{tabular}
\end{center}
\label{table:checker}
\end{table}

 %$N_C$ & $\Delta E$ for a$_{\rm Si}$ & $\Delta E$ for a$_{\rm SiGe}$ & $\Delta E^c$ for a$_{\rm Si}$ & $\Delta E^c $ for a$_{\rm SiGe}$ \\

In addition we have done DFT calculations where we considered $\rm Si_{0.5} Ge_{0.5}$ systems with alternating layers of Si and Ge.
These layers are periodic in 2 dimensions and have layer thicknesses 1 and 2. The layers are oriented along the (100) direction, and consist of a bi-layer of Si and Ge. For these systems we also find that a thicker layer is preferred by 1 meV (2 meV) when the system has the lattice constant a$_{\rm Si}$ (a$_{\rm SiGe}$). This trend continues for thicker layers (but additional changes are less pronounced).

%We have also done DFT calculations for Ag/Pt. When we stack Ag and Pt layers along the (100) direction for a system with the Ag lattice constant, we find that increasing the layer thickness from 1 to 2 (3) lowers the energy of the system by 45 (50) meV per atom. For an fcc lattice one can not construct a perfect "checkerboard" structure. But we did calculations for Ag/Pt with ``pseudo-checkerboard'' configurations, and all results (not reported here) are consistent with our assertion that more segregated systems have a lower energy.

\section{Heteroepitaxial Thin Films}
We now consider a strained flat film of thickness $T$ of
pure Ge on a semi-infinite substrate of pure Si.
This case has a free surface unlike the periodic cases discussed above.
According to continuum theory, the total elastic energy will be
\[
W=C\mu^2 A T,
\]
where $C$ depends on the elastic properties and $A$ is the area of the film.
Now suppose that Si and Ge mix perfectly in such a way that the film has a
fraction $\theta_0$ of Ge.  The film  now has a thickness $T/\theta_0$ and
an effective misfit of $\mu(\theta_0)= \theta_0\mu$~\cite{MHS}.
By continuum theory the total elastic energy is
\[
W(\theta_0)=\theta_0 W.
\]
\begin{figure}[ht]
\begin{picture}(450,100)
\centerline{\psfig{figure=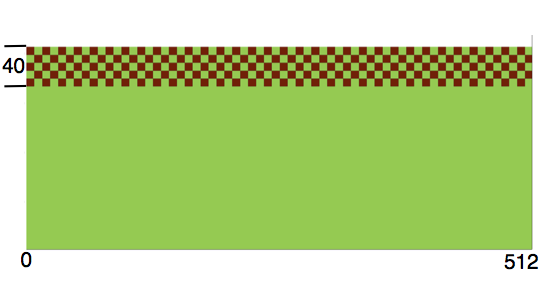,width=3.5in}}
\end{picture}
\caption{(Color Online) The eight atom checker board pattern. There are 64 ($8\times 8$) atoms
in each checker board.}
\label{CB8}
\end{figure}

\begin{table}[ht]
\caption{Normalized Total elastic energy}
\begin{center}
  \begin{tabular}{|l|c|} \hline
  1 atom checker board     &  1.241       \\[0.03in]
  \hline
  2 atom checker board     &  1.050          \\[0.03in]
  \hline
  4 atom checker board     &  0.976       \\[0.03in]
  \hline
  8 atom checker board     &  0.947        \\[0.03in]
  \hline
  random arrangement       &  1.012        \\[0.03in]
\hline
  continuum theory          &  0.500   \\[0.03in]
\hline
  \end{tabular}
\end{center}
\label{energy_table}
\end{table}
Therefore, according to continuum theory, the total elastic
energy is reduced by intermixing. In view of the previous discussion
this is not expected to be true for the ball and spring model. To that end, we
performed a numerical calculation using the method in Ref. \cite{BDS}
to calculate the elastic energy of the ball and spring with $K_L=2K_D$ and $\misfit =.04$.
The film is 40 monolayers thick and consists of an equal number of Si and Ge atoms.
We considered 4 cases in which the atoms are arranged
in checker board patterns, where the only difference is that the scale of the pattern
changes. On the finest scale, each square consists of exactly one atom.
For a reference state we computed the total
elastic energy of a thin film with 20 monolayers of pure Ge on a substrate of
pure Si. Note that the total number of Ge atoms is the same in
each of the 4 cases and in the reference state.
 We note that the reference configuration corresponds to $\theta_0 =1$
with a film thickness of 20 layers
and all the checker board patterns in the continuum model are equivalent to $\theta_0 = 0.5$
with a film thickness of 40 layers. 
The normalized elastic energy
is the total elastic energy divided by the elastic energy of the reference state.
The results are summarized in table~\ref{energy_table}.
 As noted above for the continuum theory the reference configuration has lower elastic energy.
However when microscopic segregation is accounted for in the discrete elastic model
a segregated configuration is seen to lower elastic energy over both the reference state and a random mixture of $\theta_0 =0.5$ (the continuum equivalent).
 It is in particular worth noting that the elastic energy predicted for
the continuum case is a factor of two smaller than the discrete case.
This is due to the fact that the
discrete alloys have longitudinal variations in the concentration profile and hence have
an induced strain field in the substrate that decays slowly.
This slow decay of the strain field particularly in the case of random mixtures was
observed previously in \cite{BDS} for the ball and spring model.
The homogeneous continuum case however has zero strain in the substrate leading
to a significantly lower elastic energy in comparison to the random mixture.

It is also clear that the total elastic energy significantly increases as the
length scale of the pattern decreases. This indicates that
intermixing actually increases the elastic energy for this ball
and spring model in the setting of strain thin films.

\section{Conclusions}

We have considered the behavior of the elastic energy of the binary alloys using both an atomistic model (ball and spring)
and density-functional theory. Our ball and spring calculation indicates that finely mixed alloys have
more elastic energy than those that are more coarsely mixed. This is due to the presence of microscopic strain.
The more finely mixed the alloy becomes the more difficult  it is for it to elastically relax (the system is frustrated).
The important consequence of these observations is that intermixing will actually increase the strain energy of alloy
and not lower it as predicted by continuum theories without
enthalpic contributions. One may speculate that this is an artifact of the ball and spring model but
calculations with DFT support our conclusions.

We mention that some of the difficulties faced using continuum models can be somewhat mitigated by including
an elastic component to the enthalpy of mixing. However, this would be quite challenging since we have shown
that the elastic energy of an alloy is in fact dependent on the microscopic rearrangements of the atoms
-- information typically lost in continuum models.
In many applications the atomic arrangement is constantly changing both its scale and the degree of
anisotropy, making it difficult to assess the elastic energy in terms of average values of the composition.
Simply put, the elastic energy of a material cannot be determined by the alloy
concentration alone - much more information is needed. One possibility currently being explored is the use
of something like an H-measure but with length scale information.

\section*{Acknowledgments}
We gratefully acknowledge conversations with J.M. Millunchick,  V.  Shenoy, J. Tersoff, and P. Voorhees.
This research was supported, in part, by NSF grants DMS-0810113, DMS-0854870, DMS-1115252, and DMS-0439872.

\end{document}